$SiN_x$: $Tb^{3+}$-$Yb^{3+}$, an efficient down-conversion layer compatible with silicon solar cell process

Authors

Lucile Dumont[a], Julien Cardin[a], Patrizio Benzo[a], Marzia Carrada[b], Christophe Labbé[a], Andrea L. Richard[c], David C. Ingram[c], Wojciech M. Jadwisienczak[d], Fabrice Gourbilleau[a]

[a]CIMAP CNRS/CEA/ENSICAEN/Unicaen, 6 Boulevard Maréchal Juin, 14050 Caen Cedex 4, France

[b]CEMES-CNRS, 29 rue J. Marvig, 31055 Toulouse, France

[c]Department of Physics and Astronomy, Ohio University, Athens, OH 45701, USA

[d]School of Electrical Engineering and Computer Science, Ohio University, Stocker Center, Athens OH 45701, USA

lucile.dumont@ensicaen.fr, julien.cardin@ensicaen.fr, benzo@cemes.fr, marzia.carrada@cemes.fr, christophe.labbe@ensicaen.fr, ar286106@ohio.edu, ingram@ohio.edu, jadwisie@ohio.edu, fabrice.gourbilleau@ensicaen.fr

Abstract

$Tb^{3+}$-$Yb^{3+}$ co-doped $SiN_x$ down-conversion layers compatible with silicon Photovoltaic Technology were prepared by reactive magnetron co-sputtering. Efficient sensitization of $Tb^{3+}$ ions through a $SiN_x$ host matrix and cooperative energy transfer between $Tb^{3+}$ and $Yb^{3+}$ ions were evidenced as driving mechanisms of the down-conversion process. In this paper, the film composition and microstructure are investigated alongside their optical properties, with the aim of maximizing the rare earth ions incorporation and emission efficiency. An optimized layer achieving the highest $Yb^{3+}$ emission intensity was obtained by reactive magnetron co-sputtering in a nitride rich atmosphere for 1.2 W/cm² and 0.15 W/cm² power density applied on the Tb and Yb targets, respectively. It was determined that depositing at 200 °C and annealing at 850 °C leads to comparable $Yb^{3+}$ emission intensity than depositing at 500 °C and annealing at 600 °C, which is promising for applications toward silicon solar cells.

# 1. Introduction

Silicon solar cells (SC) present an intrinsic efficiency limit calculated by Shockley and Queisser [1,2] of 29%, which is approached in the laboratory [3]. Therefore, new efficiency increasing strategies have to be investigated to further reduce the cost of the KWh. One of the promising ways of improving the efficiency that could top the actual efficiency limit is frequency conversion mechanisms. These mechanisms allow us to realize a more efficient carrier photogeneration in silicon solar cells by tuning the solar spectrum. The conversion of IR photons to visible photons with energy just above the SC band gap is called up-conversion (UC) and allows the absorption of photons by solar cells that would be lost without the UC process. This study focuses on the down-conversion (DC) process that relies on the conversion of UV photons to visible photons with an energy just above the SC band gap. This mechanism maximizes the number of photogenerated carriers with energy matching the SC band gap, at the expense of carriers with high energy. In standard solar cells, carriers having an energy largely above the solar cell band gap may transfer energy to the matrix, thus generating an energy loss mechanism that results in an overheating of the latter. This thermalization mechanism leads, therefore, to a decrease of the cell's conversion efficiency. A down-conversion (quantum cutting) process can reduce the thermalization of carriers photogenerated by absorption of photons with an energy twice the SC band gap, and can therefore increase the SC conversion efficiency. This DC process occurs in a material that absorbs a UV photon and emits two IR photons with an energy slightly above the silicon solar cell band gap. Dexter *et al.* [4] first demonstrate that this system presents a potential interest with an achievable quantum yield higher than 100%. Afterwards, many others have evidenced an internal quantum yield ranging from 120% to 199% using different matrices and rare earth (RE) ions [5–10]. Two main rare earth combinations were proposed: the co-doping using two rare earth ions and double co-doping using three [11]. Due to its greater simplicity, the first RE ions combination is more promising.

In order to obtain a DC mechanism compatible with silicon SC, two rare earth ions, $Tb^{3+}$ and $Yb^{3+}$, were chosen due to their energy configuration which allows for the down-conversion and matches the silicon band gap energy. Indeed, the transition from $^5D_4$ to $^6F_7$ energy levels of the $Tb^{3+}$ ions, corresponding to an energy of 2.53 eV (490 nm) presents twice the energy of the $^2F_{5/2}$ to $^2F_{7/2}$ transition energy of $Yb^{3+}$ ions, corresponding to an energy of 1.26 eV (980 nm). This energy configuration allows the quantum cutting or down-conversion process by the absorption of one UV photon by a $Tb^{3+}$ ion, a cooperative energy transfer (CET) to two surrounding $Yb^{3+}$ ions [12], and the emission of two IR photons (980 nm) by $Yb^{3+}$ ions. Photons emitted from $Yb^{3+}$ ions at 980 nm (1.26 eV) have an energy slightly above silicon SC band gap (1.1 eV) and are therefore absorbed with a reduce thermalization energy by a silicon solar cell. The absence of an intermediate energy level of the $Tb^{3+}$ ions prevents energy back transfer from the $Yb^{3+}$ ions, which occurs with the $Pr^{3+}$:$Yb^{3+}$ RE couple [13]. Studies of the $Tb^{3+}$:$Yb^{3+}$ couple have already been done on several matrices. Promising results were obtained in $PO_4$ phosphor and in borate glass matrices with $\eta_{QE}$=188% [12] and $\eta_{QE}$=196% [6], respectively. However those matrices presented compatibility drawbacks with the silicon SC application, such as degradation during the fabrication of solar cells. In addition, those DC layers have stability issues under highly energetic photon irradiation. Moreover, the raw materials employed may be toxic and expensive [14]. To tackle this incompatibility, our team studied $Tb^{3+}$-$Yb^{3+}$ doped $SiO_xN_y$ thin films [15] because the Si-based matrix developed is fully compatible with silicon solar cell fabrication. Moreover, this matrix presents the supplementary advantages of a higher absorption cross-section than the rare earth ions ($10^{-17}$-$10^{-18}$ cm² [16] instead of $10^{-20}$ cm² [17]) and an efficient energy transfer to the $Tb^{3+}$ ions. This system may be promising as modeling foresees an extra efficiency of 2% for a silicon solar cell topped with a $SiO_xN_y$: $Tb^{3+}$-$Yb^{3+}$ DC layer [15]. The model was developed considering a 100% CET efficiency between RE optically active ions and demonstrates the importance of incorporating as many rare earth ions as possible in the limit of RE clustering. From this standpoint and based on previous studies [18,19], we replaced the $SiO_xN_y$ matrix by the $SiN_x$ matrix in this study. Indeed, $SiN_x$ remains compatible with the silicon solar cells while being potentially able to incorporate a larger number of rare earth ions. $Si_3N_4$ layers are used on industrial solar cells as anti-reflective coatings [20], and therefore, the $SiN_x$ layer may have comparable reflectance and transmittance properties.

In this paper, the $SiN_x$: $Tb^{3+}$-$Yb^{3+}$ system is investigated with the aim of obtaining layers that have the highest $Yb^{3+}$ emission possible and might be deposited on top of a silicon solar cell. The composition and the microstructure are first investigated, followed by the optical properties. The system is optimized with the figure of merit of improving the cooperative energy transfer between $Tb^{3+}$ and $Yb^{3+}$. A comparison with the best coupling efficiency obtained in our previous study of $SiO_xN_y$:$Tb^{3+}$-$Yb^{3+}$ system will be given.

## 2. Material and methods

The samples used in this study were deposited on p-type 250 μm-thick [001] 2" silicon (Si) substrates by reactive magnetron co-sputtering in a nitrogen rich plasma. The Si/N ratio in the matrix was monitored either by the applied power density on the Si target or adjusted by the $Ar/N_2$ ratio of the gas flux injected. In order to obtain a constant matrix composition for all samples of the study, the target power density applied on the silicon was fixed at 4.5 W/cm$^2$ while the gas inlet of argon and nitrogen was fixed at 8 sccm and 2 sccm, respectively. The plasma pressure was set at 3 mTorr. Those parameters were determined according to the work of Debieu *et al.* [18] who has analyzed the influence of deposition parameters, such as the Si target power density, the chamber pressure, and the gas flux, on the matrix composition and its optical properties. The deposition temperatures were set at 200 °C and 500 °C. In order to monitor the rare earth ions incorporation, the target power density of the $Tb^{3+}$ ($RFP_{Tb}$) and $Yb^{3+}$ ($RFP_{Yb}$) was regulated independently between 0.45 and 2.4 W/cm² for the Tb target and between 0.2 and 0.9 W/cm² for the Yb target, respectively. The deposition time was adjusted to obtain 90 nm thick films. After deposition, the samples were annealed by a 1 h-classical thermal annealing (CTA) or 10 min-rapid thermal annealing (RTA) process at different temperatures ranging from 600 °C to 950 °C. In this work, with the exception of samples used in the annealing study (section 3.2.3), samples were annealed by CTA at 850 °C for 1 h.

The composition of the samples was investigated by means of Rutherford Backscattering Spectroscopy (RBS) at the Edwards Accelerator Laboratory of Ohio University using a 4.5 MV tandem accelerator. Samples, oriented with an angle of 7.5° toward beam direction, were irradiated with a 2.2 MeV $^4He^+$ ion beam. The energy resolution obtained on the sample was 20-30 keV, depending on the depth at which the analysis was performed. The elemental concentrations were calculated by fitting the data using the RUMP simulation software [21]. Stoichiometric fitting, based upon growth conditions, was completed but provided an unsatisfactory representation of the data. In order to best fit the data, non-stoichiometric fitting was completed, leading to composition errors as follows: 10% for the silicon content, 10% for the rare earth content, and 12.5% for the N content.

Fourier Transform Infrared measurements (FTIR) were performed on layers thanks to a Thermo Nicolet Nexus 750 II spectrometer working in the 4000-400 cm$^{-1}$ range, with a 5 cm$^{-1}$ resolution at room temperature. By fitting the spectra obtained with the help of simulated Gaussian functions, the vibration band positions were accessed. The silicon (Si) excess percentage of the sample was obtained according to the method of Debieu *et al*. [18] by considering the Si excess, $Si_{ex}$:

$$Si_{ex} = \frac{4-3x}{4+4x}, \quad (1)$$

with x linearly dependent on the Si-N LO peak's position, $\vartheta_{LO}(x)$, according to the equation:

$$\vartheta_{LO}(x) = 323.4\left(x - \frac{3}{4}\right) + \vartheta_{LO}\left(\frac{3}{4}\right), \quad (2)$$

with $\vartheta_{LO}\left(\frac{3}{4}\right) = 1197 \; cm^{-1}$

Ellipsometric measurements were performed to determine the complex refractive index and the thickness of the films using a UVISEL Jobin-Yvon ellipsometer with an incident angle of 66.2°. The experimental $I_c$ and $I_s$ ellipsometry spectra were recorded on a 1.5-5 eV range with 0.01 eV resolution. The determination of the refractive index n and the layer thickness, have been achieved by fitting the experimental data by a dispersion law derived from the Forouhi-Bloomer model [22] for amorphous semiconductors using the DeltaPsi2 software.

High Resolution Transmission Electron Microscopy (HREM) was carried out on a cross-sectional specimen using a Cs corrected TEM-FEG microscope Tecnai F20S.

In order to study the emission and excitation properties, photoluminescence (PL) and photoluminescence in excitation (PLE) experiments were performed at room temperature on the annealed layers using an in-house developed setup. Specifically, the PL setup included a Lot-Oriel 1 kW Xenon lamp connected to an OMNI300 monochromator used as a tunable light source. The PL spectra were recorded with a Hamamatsu (R5108) photomultiplier tube after the dispersion of the PL signal by a MSH 300 OMNI monochromator. The detection system was locked in with a SR830 amplifier referenced at the excitation light beam chopped frequency. The PL and PLE spectra obtained were corrected by the sample's thickness and by the lamp emission intensity in the PLE case.

The PL lifetimes were recorded using a time-resolved photoluminescence in-house setup. This setup is composed of a NT340 EKSPLA optical parametric oscillator (OPO) tunable source providing 5 ns FWHM pulse at different wavelength. The signal is detected by means of a THR1000 Jobin Yvon monochromator and a liquid-nitrogen-cooled Hamamatsu-NIR PMT R5509_73 InP/InGaAs detector. The decay time measurements were performed at an excitation wavelength of 244 nm with an average power of 15 mJ, and with a repetition rate of 10 Hz at room temperature. The data were acquired on a Tektronix TDS3012B oscilloscope and recorded on a computer by a Labview program.

A simulation program based on the transfer matrix method (TMM) [23] was run based on the deduced ellipsometric parameters (complex refractive index of a layer, substrate, superstrate and stack of layers, and thickness) to calculate the absorbance, reflectance, and transmittance of the fabricated DC films deposited on top of a silicon SC. Moreover, based on incoming AM1.5 solar spectral irradiance [24] on the sample surface, the solar spectral irradiance absorbed or transmitted by the DC layer has been evaluated in the considered spectral range of 300-1200 nm.

An extended transfer matrix method (ETMM) was also run in order to model the propagation of light emission in layered media [25,26]. This method leads us to the distribution of emission versus emission angle, and versus wavelength in each considered medium. The thickness and the ellipsometric parameters of the layer and surrounding medium were taken into account. The $Yb^{3+}$ spectral line shape was assumed to be described by a Lorentzian function centered at 980 nm with a FWHM of 60 nm. PL spectra were obtained by the integration over emission angles in the conversion layer, air, and silicon materials. A supplementary integration led us to the total power densities in three media leading to the global emission properties. The extraction efficiencies are obtained by defining the extracting efficiency as the ratio of the power density in a medium by the total emitted power density.

## 3. Results and discussion

In this section, the microstructure and composition of down-conversion layers will be first investigated. Their optical properties will then be studied as a function of deposition and annealing parameters. Spectral absorbance, reflectance, and luminance of a typical layer will be studied and compared to standard $Si_3N_4$ material used as an SC antireflective layer.

*3.1. Microstructure*

A FTIR study was conducted at the Brewster angle on the produced layers. Fig. 1 exhibits the transmittance FTIR spectrum of a typical $SiN_x$: $Tb^{3+}$-$Yb^{3+}$ layer. As expected, in the case of the studied fabrication process, no peaks related to the Si-H or N-H bonds between 2200 and 2090 $cm^{-1}$ and 3320 and 2500 $cm^{-1}$ were observed. Similarly, no peak corresponding to the Si-O bond around 1200 $cm^{-1}$ could be detected. These results are indicative of oxygen and hydrogen free samples. This lack of H- or O- related bands allows us to assume that the two peaks of the spectrum are only due to the Si-N bond. Those bands were fitted using two Gaussian functions peaking at 1090.8 $cm^{-1}$, and 787.9 $cm^{-1}$ shown as dashed lines on Fig. 1. Those two peaks (from left to right) are attributed to the longitudinal optical-(LO), and the transversal optical-(TO) modes of the asymmetric Si-N bond, respectively. The sample composition (x -from $SiN_x$- and Si excess) is deduced, as shown in Section 2, from the value of the LO peak position according to the procedure of Debieu *et al*. [18]. In our case, all of the samples present a Si excess ranging from 8.62 % to 20.32% (Fig. 1 (a) and (b)). The deposited and CTA annealed layers' refractive indexes were studied by ellipsometry for various compositions. Thus, the presence of oxygen in our layer would lead to a decrease of the refractive index with respect to the $Si_3N_4$ one ($n_{Si3N4}$ = 2.02 at 1.95 eV), whereas an incorporation of a Si excess would favor an increase of the refractive index (Fig. 2). The fact that all of the samples present a refractive index higher than that of $Si_3N_4$ (Fig. 2) confirms that the matrix contains a Si excess. No oxygen peak is detected on the RBS spectrum, attesting that the layer does not contain oxygen or at least a concentration lower than the detection threshold (0.2 at.%). These analyses point out the fact that our deposited layers are oxygen free and contain a Si excess.
The influence of the rare earth ions incorporation on the microstructure is then studied. For an increase of the $RFP_{Tb}$ or the $RFP_{Yb}$, the LO peak of the FTIR spectra shifts toward lower wavenumbers (Fig. 1 (a) and (b)) while the refractive index increases (Fig. 2). This might be due to the incorporation of the rare earth ions in the matrix. In this case, the increase of the internal constraints caused by adding a larger chemical element in the matrix or the increase of the defect concentration may lead to a shift as already observed in previous papers [18,27]. However, previous studies [18,28] stated that, for a non doped matrix without hydrogen or oxygen, such behavior could be caused by a change in the [N]/[Si] ratio that would come from the intrinsic modification of the Si-N bond configuration. Haseagawa *et al*. [29] have calculated that the LO peak shift is linked to the change of the length of the Si-N bond caused by a variation in the matrix composition. Thus, the FTIR and ellipsometry show an increase of the incorporation of either the silicon or RE ions, or both kind of element as the power on the Tb and Yb targets rises.

The concentrations of the different elements in the deposited layers are deduced from the RBS measurements, and are $2.8 \times 10^{17}$ atoms/cm² for Si, $4.6 \times 10^{17}$ atoms/cm² for $N_2$, $7.0 \times 10^{16}$ atoms/cm² for $Tb^{3+}$, and $1.1 \times 10^{16}$ atoms/cm² for $Yb^{3+}$ (for a sample deposited with 1.2 W/cm² on the Tb target and 0.2 W/cm² on the Yb target). Those values are converted in percentage (at.%) and compared to the results obtained for the oxynitride $SiO_xN_y$: $Tb^{3+}$-$Yb^{3+}$ layers in Ref. [15]. Only the percentage of rare earth ions are shown here: 8.5 at.% of $Tb^{3+}$ and 1.3 at.% of $Yb^{3+}$ for our layer, whereas, in the oxynitride matrix it was found 1.8 at.% of $Tb^{3+}$ and 2.0 at.% of $Yb^{3+}$ for the previous one. A microstructural study on the same $SiN_x$: $Tb^{3+}$-$Yb^{3+}$ layer using a high resolution transmission electron microscope (HRTEM) (Fig. 3) shows that the layer is homogeneous with no remarkable segregation of species (no cluster) at the observation scale. This picture coupled with the RBS concentrations evidence the larger incorporation of rare earth ions in our matrix ($SiN_x$) than in the previous one ($SiO_xN_y$) without clustering of rare earth elements.

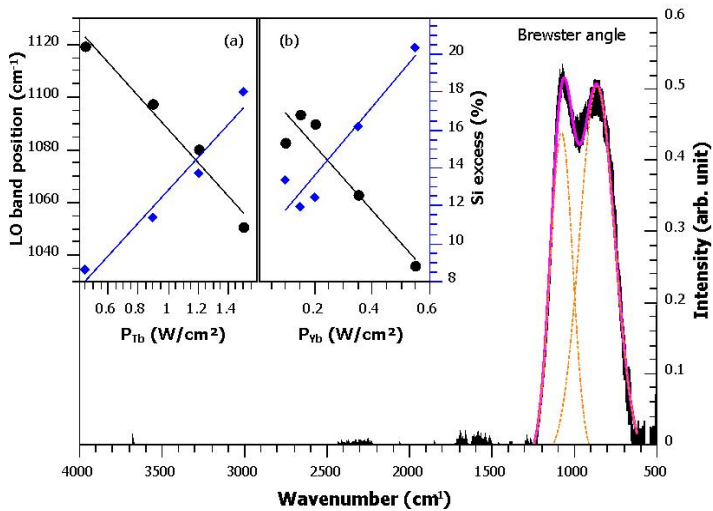

Fig. 1: **FTIR Spectra and analysis**

FTIR spectra at a 65° incidence angle (Brewster angle) of a typical $SiN_x$: $Tb^{3+}$-$Yb^{3+}$ layer. The inset presents the LO band position (black) and the value of x (blue) for samples deposited with varying $RFP_{Tb}$ for 0.2 W/cm² $RFP_{Yb}$ (a), and varying $RFP_{Yb}$ for 1.2 W/cm² $RFP_{Tb}$ (b).

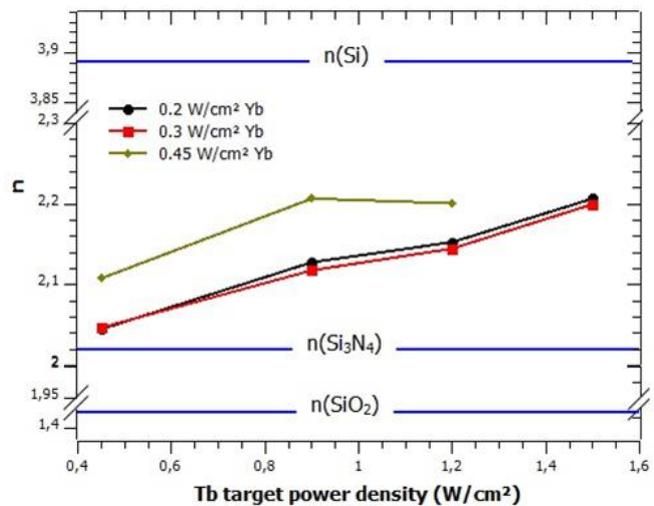

Fig. 2: **Refractive index evolution with $RFP_{RE}$**

Refractive index at 1.95 eV of $SiN_x$: $Tb^{3+}$-$Yb^{3+}$ layer (obtained by ellipsometry) according to the $RFP_{Tb}$ for different $RFP_{Yb}$. The blue line represents the refractive index of Si, $Si_3N_4$, and $SiO_2$ from top to bottom.

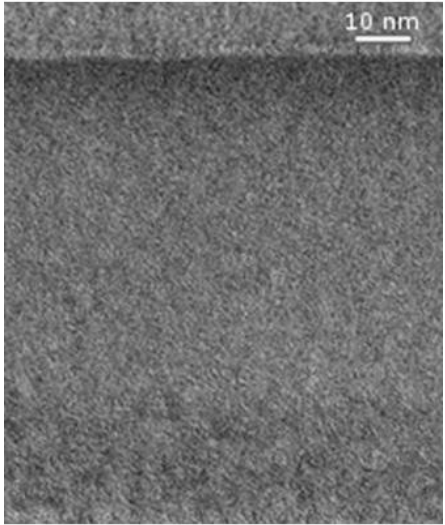

Fig. 3: **HRTEM picture**

HRTEM picture of a typical $SiN_x$: $Tb^{3+}$-$Yb^{3+}$ layer with the Si substrate at the top and the layer at the bottom.

*3.2. Down-conversion system*

In this part, we detail the down-conversion process occurring in the $SiN_x$: $Tb^{3+}$-$Yb^{3+}$ layer. The optimization process targeting the highest emission intensity (980 nm) is detailed. We also perform a comparison of absorbance, transmittance and irradiance balance between $SiN_x$: $Tb^{3+}$-$Yb^{3+}$ and $Si_3N_4$ layers on the whole spectral range (300-1200 nm) by the TMM method, considering the refractive index and thicknesses determine experimentally. Finally, we compare the $SiO_xN_y$: $Tb^{3+}$-$Yb^{3+}$ and $SiN_x$: $Tb^{3+}$-$Yb^{3+}$ samples, which exhibit the highest $Yb^{3+}$ PL emission.

*3.2.1. Down-conversion process*

Fig. 4 (main and (b)) displays the photoluminescence spectra of the $Tb^{3+}$-$Yb^{3+}$ co-doped samples for a 325 nm excitation wavelength. The inset represents the photoluminescence of the same $Tb^{3+}$-$Yb^{3+}$ layer for a more dispersive grating which allows a higher resolution of the emission peaks at the expense of the intensity. The two peaks at 990 nm and 1025 nm coming from the $^2F_{5/2}$ to $^2F_{7/2}$ transition, characteristic of $Yb^{3+}$ emission, are clearly identified.

The Yb-doped sample emits a PL signal that is slightly above the measurement noise level around 980 nm while the $Tb^{3+}$-$Yb^{3+}$ co-doped sample presents a distinct peak at 990 nm (Fig. 4). This result evidences that (i) there is no direct energy transfer between the matrix and the Ytterbium ions and (ii) Terbium ions sensitize the $Yb^{3+}$ ions under a UV excitation. A similar effect has been observed by An *et al*. in an oxynitride matrix $SiO_xN_y$ [15]. In addition, the excitation at the non-resonant wavelength of the $Tb^{3+}$ ion means that indirect excitation by the light source occurs. Thus, the $Yb^{3+}$ PL peak is a signature of double energy transfer, first from the matrix to the $Tb^{3+}$, and then the cooperative energy transfer between $Tb^{3+}$ and $Yb^{3+}$.

The PLE spectra of the co-doped system (Fig. 4 (a)) shows, after fitting the curve with Gaussian function, three peaks at 250 nm (4.96 eV), 290 nm (4.28 eV), and 330 nm (3.76 eV). We attribute those PLE peaks to the matrix band gap, the band tail, and the N-dangling defects [30–32]. From the photoluminescence results, we were able to draw an energy level scheme of the energy transfers occurring in our samples (Fig. 5). In this figure, we observe that the PLE intensity continuously decreases with the wavelength, and for wavelengths below 450 nm (2.75 eV), the PLE signal decreases to the limit of detectability (noise limit). Moreover, due to the discrete nature of the $Tb^{3+}$ energy levels and to the energy of $^5D_4$ level located at around 2.58 eV, the energy transfer from the matrix to the $Tb^{3+}$ ion may occur preferentially from the matrix UV levels to the $Tb^{3+}$ levels above $^5D_4$. Those levels with an energy above the $^5D_4$ energy level which match the energy of matrix peaks at 250 nm, 290 nm, and 330 nm could be $^4F_7$ 5d, $^5H_{6,7,8}$, and $^5L_{6,7,8}$ respectively. Energy transfer to those upper levels may be followed by a non-radiative allowed transition to the $^5D_4$ level of $Tb^{3+}$ from which cooperative energy transfer to $Yb^{3+}$ may occur.

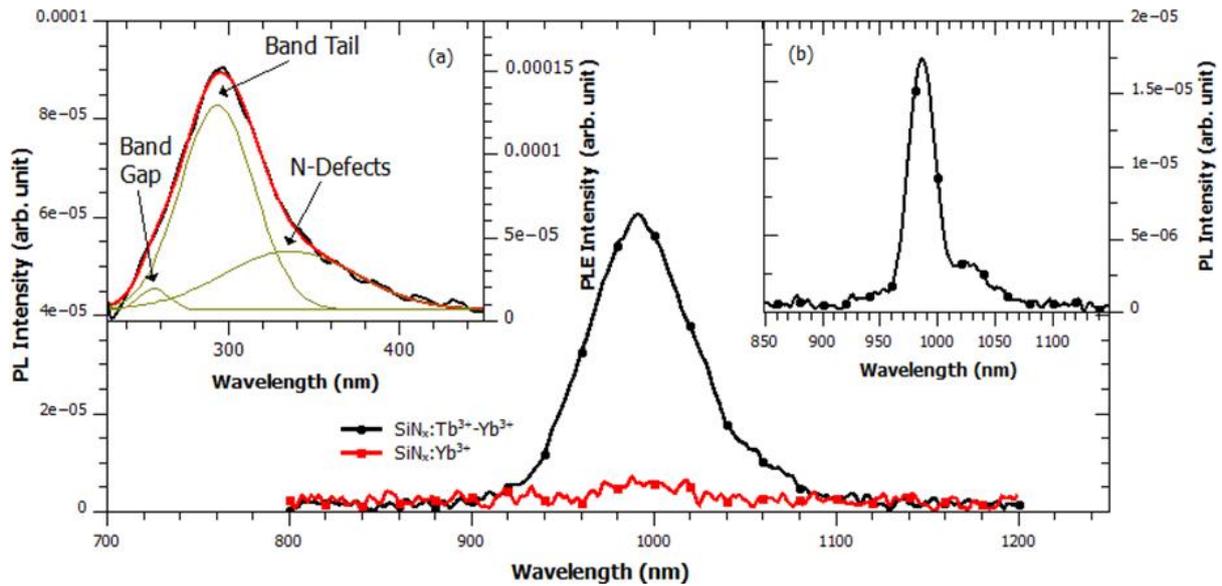

Fig. 4: **Energy transition analysis**

Photoluminescence curves of the typical co-doped and mono-doped $Yb^{3+}$ deposited layers for a non Tb resonant excitation wavelength of 325 nm measured with a grating of 600 s/mm-750 bw. The inset (a) represents the PLE spectrum of the typical $SiN_x$: $Tb^{3+}$-$Yb^{3+}$ layer (black) for a detection wavelength of 990 nm. The spectrum is fitted (red curve) with three fitting Gaussian functions (green). In the inset (b), photoluminescence curves of the typical co-doped layer measured with a grating of 1200 s/mm-300 bw are shown.

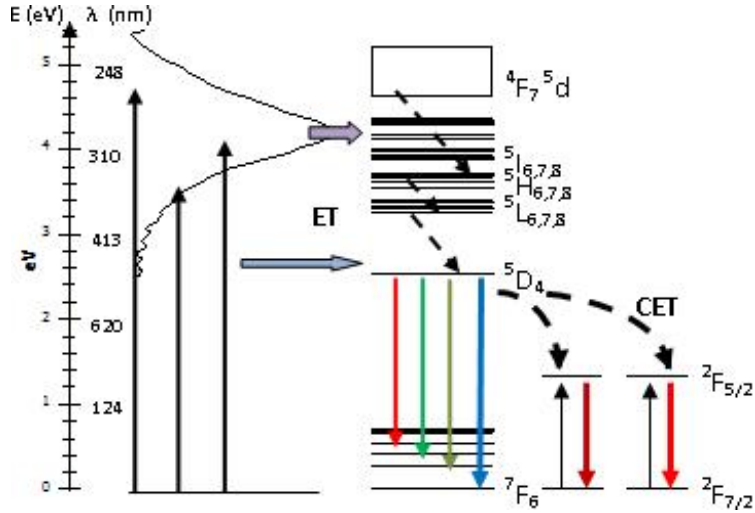

Fig. 5: **Down-conversion process**

A schematic of the energy levels of the down-conversion system with the excitation in black arrows, the radiative de-excitation (photon emission) in colored arrows, the non-radiative de-excitation in black straight dashed line arrows, the energy transfer in big blue arrows, and the cooperative energy transfer in black curved dash line arrows.

*3.2.2. Conversion efficiency optimization*

Fig. 6 presents the evolution of the $Tb^{3+}$ ($^5D_4$ to $^6F_7$ energy transition) and the $Yb^{3+}$ ($^2F_{5/2}$ to $^2F_{7/2}$ energy transition) PL peak intensity for an increasing $RFP_{Yb}$. The 545 nm-$Tb^{3+}$ intensity drops to the noise value of the measurement as soon as $Yb^{3+}$ ions are incorporated in the film, suggesting that an efficient cooperative energy transfer takes place between $Tb^{3+}$ and $Yb^{3+}$. Because the $Tb^{3+}$ peak intensity is indiscernible from the noise level, we will focus on the 990 nm-$Yb^{3+}$ PL peak emission in the following study. It decreases with the rise of the $RFP_{Yb}$, consequently with the increase of the $Yb^{3+}$ incorporation. This evolution in the matrix might be explained by an increase of the number of non-radiative centers due to the increasing content of RE ions, and/or by an increase of the energy migration process taking place between $Yb^{3+}$ ions that became closer to each other.

Fig. 7, shows that, whatever the power density used on the Yb target, the 990 nm-$Yb^{3+}$ maximum intensity increases with $RFP_{Tb}$ up to 1.2 W/cm² before decreasing. This feature is the signature that higher $Tb^{3+}$ ion concentration is required to efficiently sensitize the $Yb^{3+}$. However, a larger $Tb^{3+}$ incorporation may lead to a larger disorder of the matrix. On the one hand, one can assume that for $RFP_{Tb}>1.2$ W/cm² the PL efficiency is maximum, while adding more $Tb^{3+}$ ions only increases the disorder of the matrix, creating more non radiative de-excitation channels and thus decreasing the PL emission intensity. On the other hand, an increasing number of $Tb^{3+}$ ions may favor the energy migration process due to decreasing $Tb^{3+}$-$Tb^{3+}$ distance which increases the probability of transport of the excitation energy to a defect, where non radiative decay may occur and can therefore reduce the global emission intensity.

The increase of the $RFP_{Yb}$ also has a detrimental effect on the 990 nm-$Yb^{3+}$ emission intensity (Fig. 6 and Fig. 7(a)). This drop might be explained by the same hypothesis detailed above with the increase of the disorder of the matrix and/or an energy migration process between $Yb^{3+}$ ions. The effect of lower $RFP_{Yb}$ on $Yb^{3+}$ emission was then investigated. Fig. 7(b) shows the photoluminescence spectra of samples fabricated with 1.2 W/cm² $RFP_{Tb}$ and different $RFP_{Yb}$ ranging from 0.1 to 0.55 W/cm². An increase of the photoluminescence intensity for $RFP_{Yb}$ up to 0.15 W/cm² followed by a drop is observed. This evolution shows the higher sputtering yield of the Yb species compared to Tb, and demonstrates the critical role of the incorporation of $Yb^{3+}$ ions in such doped layers. Before 0.15 W/cm² $RFP_{Yb}$, we suppose that the number of optically active $Yb^{3+}$ ions increases with the addition of $Yb^{3+}$ ions, and thus, favors the $Yb^{3+}$ emission intensity. After reaching 0.15 W/cm² for $RFP_{Yb}$, the drop in the emission intensity may be attributed to the incorporation of optically inactive $Yb^{3+}$ ions or to an increase of defects density. So, deduced from Figs. 6 and 7, the optimized sample is obtained for a RF power density of 1.2 W/cm² and 0.15 W/cm² applied on the Tb and Yb targets, respectively.

Fig. 1 shows that the silicon excess $Si_{ex}$, increases with the $RFP_{Tb}$ and $RFP_{Yb}$. In the meantime, the PL intensity of the $Yb^{3+}$ peak increases to an optimum before decreasing. Thus no direct link between this Si excess and the emission mechanisms (the PL intensity) was evidenced.

In order to investigate the cause of the $Yb^{3+}$ intensity drop, lifetime experiments were carried out on samples with varying power density on the Yb target (Fig. 8). The lifetime curves present a nonsingular exponential behavior, and therefore we use a mean decay time model $\tau_m$ given by :

$$\tau_m = \int_0^\infty \left(\frac{I(t)}{I_0}\right) dt. \tag{3}$$

We observe that the $Yb^{3+}$ lifetime drops while $RFP_{Yb}$ increases up to 0.45 W/cm². Above this power density threshold, the lifetime drop characterizes the presence of supplementary non-radiative de-excitation channels such as defects or migration processes between $Yb^{3+}$ ions.

The RBS measurements, as well as the HRTEM observations, (Fig. 3) were performed on the best sample ($SiN_x$-1.2Tb-0.15Yb yielding high PL). The most intense emission is obtained for a layer containing a high concentration of rare earth ions (8.5 at.% of $Tb^{3+}$ and 1.3 at.% of $Yb^{3+}$) without any evidence of RE clusters formation. In Fig. 9, the PL intensity at 990 nm of our more luminescent sample ($SiN_x$-1.2Tb-0.15Yb) is found to be 6 times more intense than the more luminescent sample obtained on the $SiO_yN_y$: $Tb^{3+}$-$Yb^{3+}$ system [15]. It has been shown [33] that the $Yb^{3+}$ PL intensity, $I_{PL}$, is given by:

$$I_{PL} = \frac{N^* \cdot V}{\tau_{rad}}, \tag{4}$$

with V being the volume on which the PL is applied, $N^*$ being the excited state population density, and $\tau_{rad}$ being the $Yb^{3+}$ radiative lifetime. In our experiment, the volume of interaction and the pump power density ($1.9 \times 10^{15}$ photons/cm² at 285 nm) are constant. Assuming a constant $\tau_{rad}$ and using Eq. (5), we obtain the $Yb^{3+}$ excited state ratio $N^*_{SiN_x}(Yb^{3+})/N^*_{SiO_xN_y}(Yb^{3+}) = 6$, that demonstrate the higher PL efficiency of the $SiN_x$: $Tb^{3+}$-$Yb^{3+}$ system.

For all the studies presented here, the 1 h-CTA 850 °C annealing was applied; however, a possible degradation by high temperature annealing has to be taken into account in the fabrication process of photovoltaic cells. Indeed, above 700 °C the P and B dopant diffusion increases drastically destroying the PN-junction. Therefore new set of layers have been deposited at 500 °C followed by CTA or RTA annealing treatments at different temperatures. In Fig. 10, we observe that depositing at 500 °C favors the increase of the photoluminescence intensity compared to depositing at 200 °C, thus allowing the use of a lower annealing temperature for a similar result. Moreover, the RTA process leads to slightly higher emission intensity with a lower temperature and/or duration (lower thermal budget) which limits the diffusion of the species in SC. The same intensity of photoluminescence as with samples deposited at 200 °C and annealed with 1 h-CTA 850 °C was therefore obtained with samples deposited at 500 °C and annealed with 10 min-RTA 600°C, allowing the use of our DC layer in the process of solar cell fabrication.

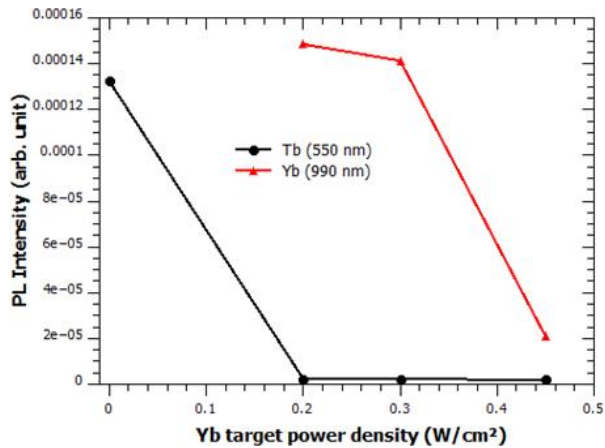

Fig. 6: **PL peak intensity**

$Tb^{3+}$ and $Yb^{3+}$ peak intensity as a function of the $RFP_{Yb}$ for an excitation wavelength of 285 nm measured with a grating of 600 s/mm-750 bw.

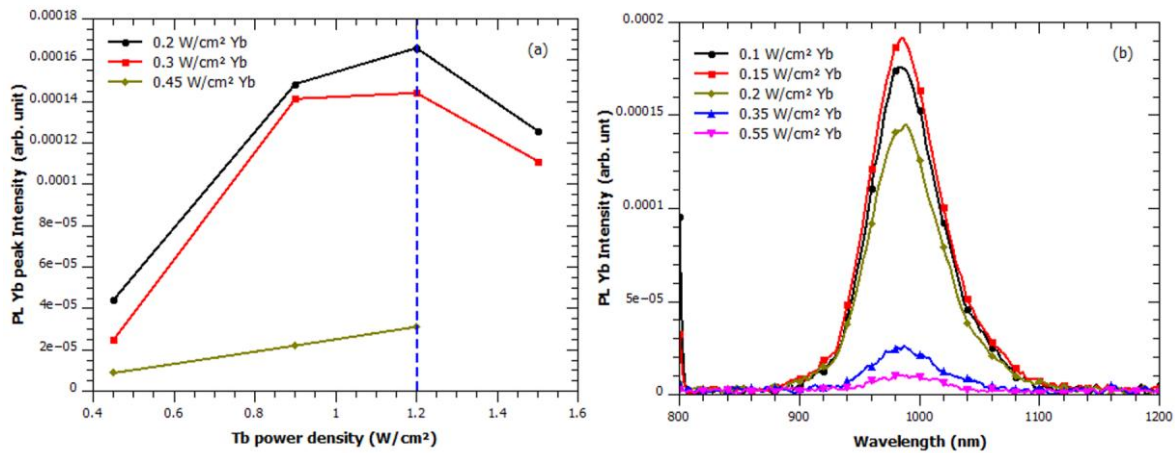

Fig. 7: **Evolution of $Yb^{3+}$ emission versus $RFP_{RE}$**

(a) Yb3+ peak intensity as a function of the $RFP_{Tb}$ for different $RFP_{Yb}$ and for an excitation wavelength of 285 nm measured with a grating of 600 s/mm-750 bw; blue dashed line at 1.2 W/cm² is for the Tb target.

(b) Photoluminescence spectra of co-doped samples with various $RFP_{Yb}$ for a 1.2 W/cm² $RFP_{Tb}$ and for an excitation wavelength of 285 nm measured with a grating of 600 s/mm-750 bw.

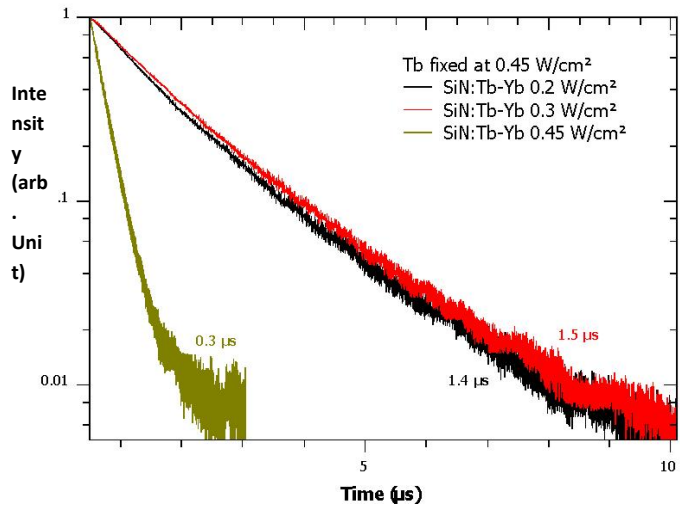

Fig. 8: **Yb$^{3+}$ decay time versus RFP$_{Yb}$**

Time-resolved photoluminescence of Yb$^{3+}$ for different RFP$_{Yb}$ for an excitation wavelength of 244 nm. The lifetimes corresponding to each curve are given in the same color.

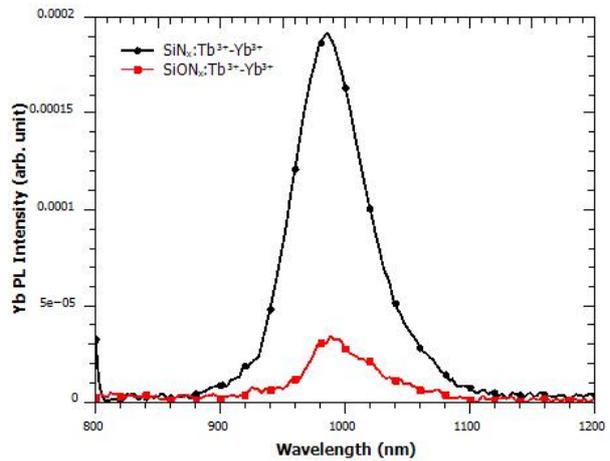

Fig. 9: **Yb$^{3+}$ emission for different matrixes**

Photoluminescence spectra of the best samples obtained of SiN$_x$ and SiO$_x$N$_y$ co-doped matrixes for an excitation wavelength of 285 nm measured with a grating of 600 s/mm-750 bw.

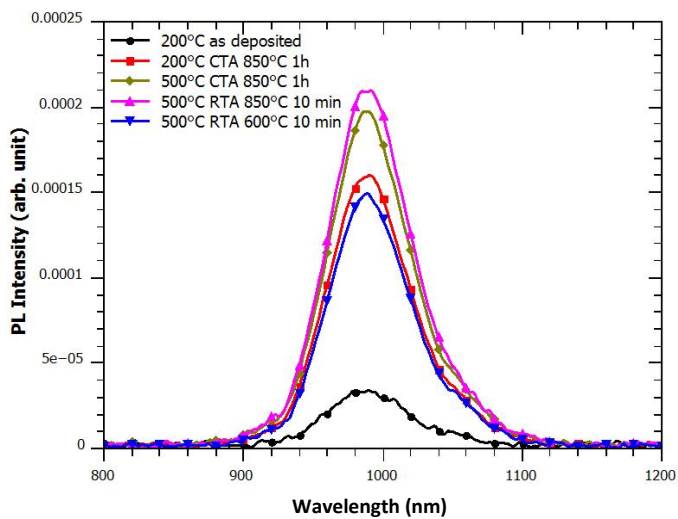

Fig. 10: **Yb$^{3+}$ emission for different annealing processes**

Photoluminescence spectra of the standard sample for different annealing processes for an excitation wavelength of 285 nm measured with a grating of 600 s/mm-750 bw.

*3.2.3. Simulation of the absorption properties*

In addition, to optimize a fabrication process compatible with the silicon solar cells, it is important to determine the impact that adding such layers will have on the light received and converted by the solar cell. For this purpose, the optical properties of our layer are studied.

Following the ellipsometry measurement, we developed a TMM program that enables us to calculate/access the absorbance (A), reflectance (R), and spectral irradiance that was incident, transmitted and absorbed by the layer. Our goal is to compare the results of the deposited layer to those of a standard $Si_3N_4$, anti-reflective coating layer used commonly on silicon solar cells. The previous variables will be integrated over the spectral range and normalized by the integration of the spectral range.

It was found that the different deposited samples studied (for different power densities on the Tb and Yb targets) in this paper have comparable values of integrated absorbance, reflectance, and irradiance. Thus, the quantity of rare earth incorporated does not influence those optical properties significantly (the maximum change is about 1%). When compared to a standard $Si_3N_4$ anti-reflective coating layer on Fig. 11, one can see that our layer's reflectance is comparable to the $Si_3N_4$ one (14% integrated reflectance for both). Our layer has thus the same anti-reflective properties as the $Si_3N_4$. The integrated absorbance of our layer is stronger (ten times), and so it is useful in the UV part of the spectrum as it allows a better use of the UV photons, but it is less useful in the visible part of the spectrum as it absorbs photons that could be used by the solar cell below. However, the integrated radiances out of the layer for our samples or for the $Si_3N_4$ are similar (79% and 85%) pointing that our layers do not have a negative impact on the light received by the solar cell immediately after the down-conversion layer. The similarity of the optical properties between our sample and the $Si_3N_4$ ensures that our layers can be topped on silicon solar cell with a good prospect of efficiency improvement.

In the same way, the normalized absorbance of both $SiO_xN_y$: $Tb^{3+}$-$Yb^{3+}$ and $SiN_x$: $Tb^{3+}$-$Yb^{3+}$ samples exhibiting the highest $Yb^{3+}$ PL emission are estimated to be 17.9% and 5.35%, respectively, for an excitation at 285 nm and $1.9\times10^{15}$ photons/cm². This difference can be attributed to opto-geometrical differences between the two studied layers, where the layers have different thicknesses 225 nm and 90 nm and different refractive indexes. It is worth noticing that the same values are obtained on the UV part of the spectrum and on the whole spectrum as well. This means that more UV photons are absorbed in the $SiO_xN_y$ matrix than in the $SiN_x$ matrix, but also that the IR photons emitted by the DC process are also more absorbed and thus less transmitted to the solar cell. It is confirmed by the use of the ETMM program which yields the extraction efficiencies in the air, in the layer, and in the substrate. The extraction efficiencies are the ratio of the emission in the medium we seek (air, layer or substrate) over the total emission, they are given in Tab. 1. The $SiN_x$ matrix transfers more to the SC (modeled here by the Si) than the $SiO_xN_y$ matrix.

For both samples, assuming that their PL intensity is proportional to a global efficiency, $\eta^{glo}$, composed of the absorbance, the extraction, and the internal efficiencies:

$$I_{PL} \propto \eta^{glo} = \eta^{abs}\eta^{int}\eta^{ext} \ . \tag{5}$$

The extraction efficiencies, $\eta^{ext}$, defined for an extraction medium are given in Table 1. The absorbance efficiencies correspond to the normalized absorbances: $\eta^{abs}(SiN_x)$=5.36% and

$\eta^{abs}(SiN_xO_y)=17.9\%$. The internal efficiency represents both the DC conversion efficiency as well as the propagation with partial absorption of the signal at 980nm ($Yb^{3+}$) to the interfaces of the emitting medium. As previously mentioned in section 3.2.2, the $Yb^{3+}$ excited state ratio of $SiN_x$ DC layer (Fig. 9) was found to be 6 times higher than the one for $SiN_xO_y$ layers with the same PL experimental conditions ($1.9\times10^{15}$ photons/cm² at 285nm). From this ratio and with relationship (Eq. (5)) we determined the internal efficiency ratio:

$$\eta^{int}(SiN_x)/\eta^{int}(SiN_xO_y) = 28.9.$$

This internal efficiency improvement of the DC $SiN_x$ layer compared to the thicker $SiN_xO_y$ can be attributed to following improvements: (i) the optically active RE content, (ii) the DC conversion, and (iii) the 980 nm photons propagation.

The PL intensity represents the global efficiency in the air however in the silicon solar cell application, we may be interested in the global efficiency in the silicon SC. In order to evaluate this global efficiency in silicon, we use the Eq. (5) replacing the extraction ratio in the air by the extraction ratio in the silicon (Table 1). The ratio of global efficiency of DC layers with emission in the Si substrate leads to:

$$\eta^{glo}(SiN_x)/\eta^{glo}(SiN_xO_y) = 11.6. \tag{7}$$

This ratio shows the improvement of the global efficiency of the DC $SiN_x$ layer compared to the thicker $SiN_xO_y$ one in the silicon (assumed to be representative of the silicon SC). For the conditions of thickness and refractive index employed in this work, the silicon, and thus the silicon solar cells, would receive 11.6 times more light at 980 nm from the DC $SiN_x$ layer compared to the thicker $SiN_xO_y$ one.

| $\eta^{ext}$(%) | Air | Layer | Silicon |
|---|---|---|---|
| $SiN_x$ | 15.8 | 60.8 | 23.4 |
| $SiN_xO_y$ | 22.8 | 59.7 | 17.5 |

Tab. 1: **Extraction efficiencies**

Extraction efficiencies of the $SiN_x$ and $SiO_xN_y$ matrixes in the air, the layer, and the silicon.

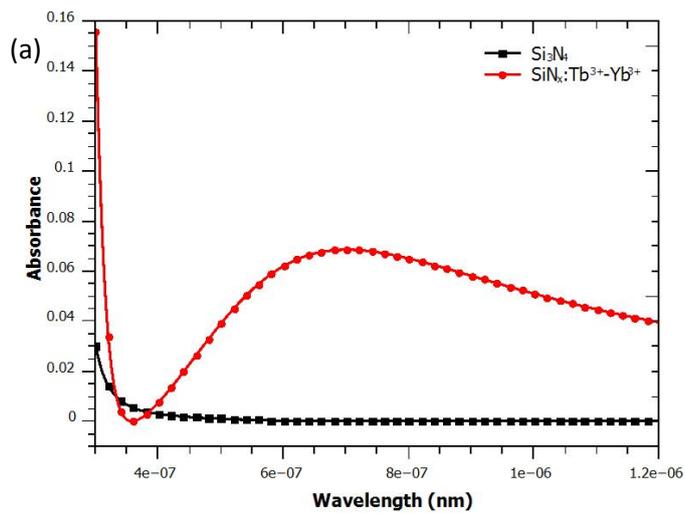

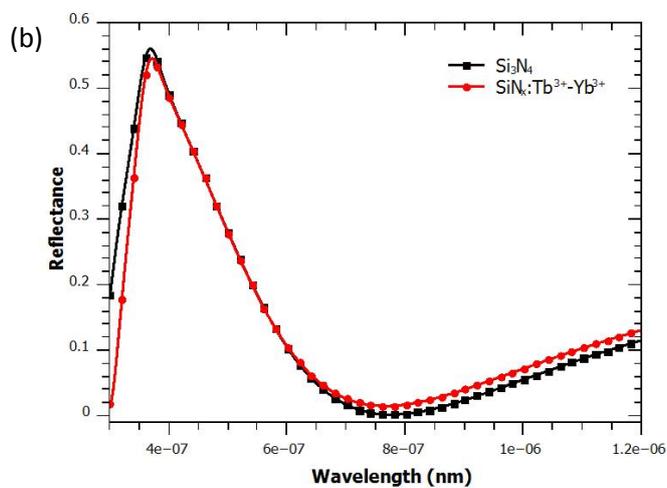

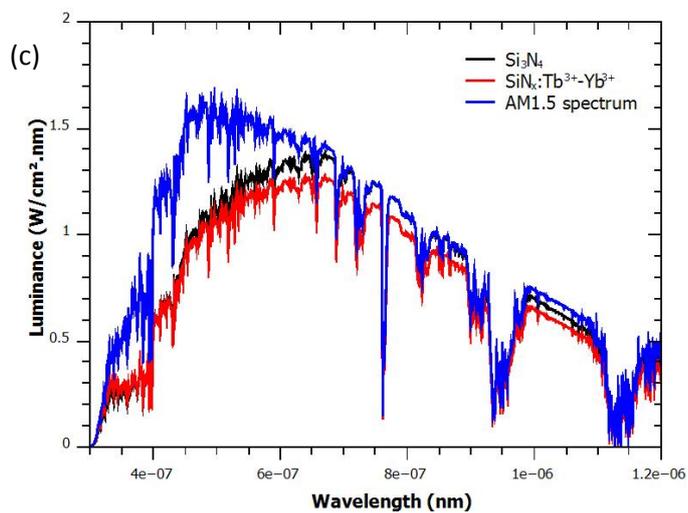

Fig. 11: **Simulated parameters**

Absorbance (a), reflectance (b), and irradiance that go out of the deposited layer (c) for the $Si_3N_4$ (black) and our typical deposited layer (red). The AM1.5 solar spectrum is represented in blue in (c).

## 4. Conclusion

In this work, we have studied the cooperative energy transfer between $Tb^{3+}$ and $Yb^{3+}$ in the $SiN_x$ matrix. The study of the composition, the microstructure, and the optical properties of the samples allows for a better understanding of the process. An optimized layer was deposited by reactive magnetron co- sputtering in a nitride rich atmosphere for a 1.2 W/cm² and 0.15 W/cm² power density applied on the Tb and Yb targets, respectively. The annealing process was also studied and leads to the conclusion that depositing at 500°C and annealing at 600°C with a 10 min-RTA process leads to the higher $Yb^{3+}$ PL emission that can be deposited on the silicon SC without damaging it. The layer obtained was compared to the more efficient sample produced during a previous study on $SiO_xN_y$: $Tb^{3+}$-$Yb^{3+}$, and found to have a 6 times higher PL efficiency for a photon flux of $1.9 \times 10^{15}$ photons/cm² at 285 nm. A 28.9 times higher internal efficiency corresponding to the DC conversion and propagation of the signal at 980nm ($Yb^{3+}$) efficiencies was found. Moreover, an efficiency of the DC layer emission into the silicon was 11.6 times higher than the one with $SiO_xN_y$: $Tb^{3+}$-$Yb^{3+}$. Considering the similarity of the Silicon substrate and a silicon solar cell material, the latter efficiency increase is promising for future applications on this DC layer stacked over a real silicon solar cell. In addition, the optical properties of our layer were found to be comparable to the ones of a $Si_3N_4$ classical anti-reflective coating, meaning that adding the layer will not inhibit the solar cell efficiency while presenting the additional efficiency provided by the DC layer. Thus, our system may enhance the silicon solar cells' efficiency by a few percentage points.

**Acknowledgement**

The authors are grateful to the French National Research Agency that provided support for this work through the GENESE Project (ANR-13-BSS09-0020-01), as well as to the CNRS and the Basse-Normandie region that supports PhD students, and the French national network METSA (www.metsa.fr) for providing access to the Tecnai microscope at CEMES laboratory. W.M.J. acknowledges the support from the NSF CAREER Award No. DMR-1056493.